\documentclass[aps,prl,twocolumn,superscriptaddress,showpacs,amsmath,preprintnumbers]{revtex4}

\usepackage{bm}
\usepackage{graphicx}
\usepackage{slashed}

\newcommand\vek[1]{\bm{#1}}
\newcommand\adj[1]{\overline{#1}}
\newcommand\LL[1]{#1_{\text{L}}}
\newcommand\RR[1]{#1_{\text{R}}}
\newcommand\imag{\mathrm{i}}
\newcommand\cc{\text{c.c.}}
\newcommand\dirac{\bm{D}}
\newcommand\Kdirac{\bm{K}}
\DeclareMathOperator{\Tr}{Tr}
\DeclareMathOperator{\tr}{tr}
\DeclareMathOperator{\re}{Re}
\DeclareMathOperator{\im}{Im}

\begin{document}

\title{Temperature Dependence of Standard Model CP Violation}

\author{Tom\'{a}\v{s} Brauner}
\affiliation{Faculty of Physics, University of Bielefeld, D-33615 Bielefeld, Germany}
\affiliation{Department of Theoretical Physics, Nuclear Physics Institute ASCR, 25068 \v{R}e\v{z}, Czech Republic}
\author{Olli Taanila}
\affiliation{Faculty of Physics, University of Bielefeld, D-33615 Bielefeld, Germany}
\affiliation{Helsinki Institute of Physics, P.O.~Box 64, FI-00014 University of Helsinki, Finland}
\author{Anders Tranberg}
\affiliation{Niels Bohr International Academy, Niels Bohr Institute and Discovery Center, Blegdamsvej 17, 2100 Copenhagen, Denmark}
\author{Aleksi Vuorinen}
\affiliation{Faculty of Physics, University of Bielefeld, D-33615 Bielefeld, Germany}

\begin{abstract}
We analyze the temperature dependence of CP violation effects in the Standard Model by determining the effective action of its bosonic fields, obtained after integrating out the fermions from the theory and performing a covariant gradient expansion. We find non-vanishing CP violating terms starting at the sixth order of the expansion, 
albeit only in the C odd/P even sector, with coefficients that depend on quark masses, CKM matrix elements, temperature and the magnitude of the Higgs field. The CP violating effects are observed to decrease rapidly with temperature, which has important implications for the generation of a matter-antimatter asymmetry in the early Universe. Our results suggest that the cold electroweak baryogenesis scenario may be viable within the Standard Model, provided the electroweak transition temperature is at most of order $1\text{ GeV}$.
\end{abstract}


%\preprint{BI-TP 2011/40}

\maketitle

%%%%%%%%%%%%%%%%%%%%%%%%%%%%%%%%%%%%%%%%%%%%%%%%%%%%%%%%%%%%%%%%%%%%%%%%

\emph{Introduction}.---One of the most fundamental observations in modern cosmology is the existence of an asymmetry between matter and antimatter in the Universe. In order to generate such a state, high energy processes must have been efficient in breaking the baryon number, C (charge conjugation) and CP (combined charge conjugation and parity) symmetries in an environment out of thermal equilibrium \cite{sakharov}. All these ingredients are present in the Minimal Standard Model (SM), but the precise mechanism to generate the correct abundance (if one exists) has so far eluded us \cite{EWBG}.

Within the SM, CP violation originates from the complex phase of the Cabibbo-Kobayashi-Maskawa (CKM) matrix. One approach to include its effects in out-of-equilibrium field dynamics is to integrate the fermions out from the theory and recover CP violating bosonic operators in the resulting effective action---a technique particularly well suited for classical numerical simulations. This was first carried out in a perturbative setting by Shaposhnikov \emph{et al.}~\cite{shaposhnikov}, who argued that at temperatures~$T$ above the electroweak scale, CP violating operators are suppressed by coefficients proportional to $T^{-12}$. The conclusion was that the SM electroweak phase transition, taking place at $T\simeq100\text{ GeV}$, cannot constitute a viable mechanism for baryogenesis (see also Ref.~\cite{gavela}).

Somewhat later, it was argued that electroweak baryogenesis might still work if there was a mechanism to delay the transition until a temperature scale of $T\simeq1\text{--}10\text{ GeV}$, a scenario typically referred to as cold electroweak baryogenesis \cite{cewbag1,cewbag2,cewbag3,cewbag4}. Refs.~\cite{smit,schmidt,salcedo6,salcedo8} subsequently employed a covariant gradient expansion to determine the leading CP violating terms in the bosonic SM effective action at zero temperature; however, with contradictory results. In particular, while Ref.~\cite{schmidt} claimed the existence of a C even/P odd operator at the sixth order of the expansion, Ref.~\cite{salcedo6} reported that only the C odd/P even sector gives a non-vanishing CP violating contribution. Following these works, numerical simulations employing the zero temperature CP violating operator of Ref.~\cite{schmidt} demonstrated that one obtains a baryon asymmetry exceeding the observed one by four orders of magnitude~\cite{tranberglong}. Taking into account the known suppression of CP violation in the limit of very high temperatures \cite{shaposhnikov}, this highlights the importance of determining the temperature dependence of CP violating effects within the SM.

In the present Letter, our objective is simple: To evaluate the CP violating part of the SM bosonic effective action as a function of temperature. The resulting rate of temperature dependence should, in combination with the results of Ref.~\cite{tranberglong}, give at least a rough estimate of the temperature range, in which SM cold electroweak baryogenesis might be viable. In the limit of zero temperature, our calculation should also provide an independent check of the very non-trivial predictions of Ref.~\cite{salcedo6}, addressing the discrepancy between the results of Refs.~\cite{schmidt,salcedo6}.

%%%%%%%%%%%%%%%%%%%%%%%%%%%%%%%%%%%%%%%%%%%%%%%%%%%%%%%%%%%%%%%%%%%%%%%%

\emph{Bosonic effective action}.---Consider a Euclidean field theory of chiral fermions, denoted collectively as $\psi(x)\equiv(\LL\psi(x),\RR\psi(x))$, coupled to gauge fields $\LL V(x),\RR V(x)$ and possibly to additional scalar fields. The dynamics of the fermions is described by a Euclidean Lagrangian of the Dirac type, $\adj\psi(x)\dirac(x)\psi(x)$, where the Dirac operator~$\dirac(x)$ in the chiral (left-right) basis has the generic form
\begin{equation}
\dirac(x)=\begin{pmatrix}
\slashed\partial+\LL{\slashed V}(x) & m_{\text{LR}}(x)\\
m_{\text{RL}}(x) & \slashed\partial+\RR{\slashed V}(x)
\end{pmatrix},
\end{equation}
using the Feynman slash notation, $\slashed v\equiv v_\mu\gamma_\mu$. The functional integral over the fermionic fields can be performed exactly, resulting in an additional contribution to the effective action of the bosonic fields,
\begin{equation}
\Gamma[\LL V,\RR V,m_{\text{LR}},m_{\text{RL}}]=-\Tr\log\dirac.
\end{equation}

As demonstrated in Refs.~\cite{salcedo1}, the P even and P odd parts of this effective action, $\Gamma^+$ and $\Gamma^-$, can be expressed in the form
\begin{eqnarray}
\Gamma^+&=&-\frac12\re\Tr(\log\Kdirac),\label{gammaplus}\\
\Gamma^-&=&-\frac12\imag\im\Tr(\gamma_5\log\Kdirac)+\Gamma_{\text{gWZW}},
\label{gammaminus}
\end{eqnarray}
where the second order differential operator $\Kdirac$ is defined as $\Kdirac\equiv m_{\text{LR}}m_{\text{RL}}-\slashed\dirac_{\text{LL}}m_{\text{RL}}^{-1}\slashed\dirac_{\text{RR}}m_{\text{RL}}$. The gauged Wess-Zumino-Witten term $\Gamma_{\text{gWZW}}$ encodes the chiral anomaly.

The trace-logarithm of the Dirac operator cannot be evaluated in a closed form for general coordinate-dependent background fields. Therefore, one usually resorts to a covariant gradient expansion, that is, an expansion in gauge fields and covariant derivatives. A useful tool for this is the method of covariant symbols~\cite{Pletnev:1998yu,salcedo1}---a general method for the evaluation of traces of differential operators of the type $f(D,m)$, where $f$ is an algebraic function of its two arguments,  a (matrix covariant) derivative $D$ and a (matrix) field $m(x)$. At nonzero temperature, the trace of this operator reads
\begin{eqnarray}
&&\Tr f(D,m)=\int_{x,p}\tr\bigl[f(\overline{D},\adj m)\openone\bigr]\label{metcovsymb}\\
&&\qquad +\int_{x,p}\sum_{k=1}^\infty\frac{(-\imag)^k}{k!}\tr\bigl[(D_0\nabla_0)^kf(\adj D,\adj m)\openone\bigr], \nonumber
\end{eqnarray}
where $\nabla_\mu\equiv\partial/\partial p_\mu$, ``$\tr$'' stands for a trace over matrix indices, and the bar denotes a conjugation, $\adj v\equiv e^{\imag D\cdot\nabla}e^{-\imag p\cdot x}v\,e^{\imag p\cdot x}e^{-\imag D\cdot\nabla}$ \footnote{Upon completion of our work, we became aware of this result having been independently obtained in Ref.~\cite{salcedonew}.}. Finally, the integration measures in a $d$-dimensional Euclidean spacetime have the forms $\int_{x}\equiv\int_0^{1/T}dx_0\int d^{d-1}\vek x$ and $\int_p\equiv T\sum_{p_0}\int d^{d-1}\vek p/(2\pi)^{d-1}$, where $p_0$ takes the values $2n\pi T$ or $(2n+1)\pi T$ with integer $n$, depending on whether the trace is taken over a~bosonic or fermionic space.

A detailed derivation of Eq.~(\ref{metcovsymb}) will be given elsewhere~\cite{longpaper}; here, we simply remark that at zero temperature, the derivative $D$ as well as the field $m$ are Lorentz covariant, the first term of this expression is completely Lorentz invariant, and the second term vanishes. At nonzero temperature this is no longer the case. One should additionally note that due to the presence of explicit free temporal derivatives in the second term, the individual terms in the sum are not gauge invariant.

%%%%%%%%%%%%%%%%%%%%%%%%%%%%%%%%%%%%%%%%%%%%%%%%%%%%%%%%%%%%%%%%%%%%%%%%

\emph{Application to the Standard Model}.---We need not write down the full SM Lagrangian, but simply display its Euclidean Dirac operator in order to fix our conventions, following closely the notation of Ref.~\cite{salcedo6}. Upon diagonalizing the mass matrices, the chiral components of the Dirac operator in the quark sector take the form
\begin{eqnarray}
\slashed\dirac_{\text{LL}}&=&\begin{pmatrix}
\slashed D_u+\slashed Z+\slashed G & \slashed W^+V\\
\slashed W^-V^{-1} & \slashed D_d-\slashed Z+\slashed G
\end{pmatrix},\label{dirac1}\\
\slashed\dirac_{\text{RR}}&=&\begin{pmatrix}
\slashed D_u+\slashed G & 0\\
0 & \slashed D_d+\slashed G
\end{pmatrix}.
\label{dirac2}
\end{eqnarray}
Here, $Z_\mu$, $W_\mu^\pm$ and $G_\mu$ are the weak intermediate boson and gluon fields, respectively, and $V$ is the CKM matrix. Furthermore, $D_{u\mu}=\partial_\mu+(2/3)B_\mu$ and $D_{d\mu}=\partial_\mu-(1/3)B_\mu$, where $B_\mu$ is the hypercharge gauge field. (Note that we are using a ``mixed'' basis in terms of $Z_{\mu}$ and the hypercharge field $B_{\mu}$ rather than the commonly used $Z_\mu$ and the photon field $A_\mu$.) The indicated $2\times2$ matrix structure of the Dirac operator corresponds to weak isospin; Dirac, color and family indices are suppressed. The left-right and right-left components of the Dirac operator are identical, $m_{\text{LR}}=m_{\text{RL}}=(\phi/v)\,\text{diag}(m_u,m_d)$, where $m_{u,d}$ are the (diagonal) mass matrices for $u$ and $d$ type quarks. Throughout the calculation, we use the unitary gauge, in which the only physical scalar field of the SM---the Higgs field $\phi$---fluctuates around its vacuum expectation value, $v\approx246\text{ GeV}$.

The above expressions for the Dirac operator completely fix our conventions. Our fields are related to the canonically normalized (Euclidean) fields, denoted here by a tilde, via $W_\mu^\pm=(g/\sqrt2)\tilde W_\mu^\pm$, $Z_\mu=[g/(2\cos\theta_{\text W})]\tilde Z_\mu$, $B_\mu=g'\tilde B_\mu$, and $G_\mu=(g_s/2)\lambda_a\tilde G_{a\mu}$. Here, as usual, $g,g',g_s$ denote respectively the weak isospin, hypercharge and strong coupling constants, while $\theta_{\text W}$ is the Weinberg angle and $\lambda_a$ are the Gell-Mann matrices.

The elements of the CKM matrix are known to a good accuracy~\cite{PDG} and are fixed up to independent rephasings of the quark fields, which must leave all physical observables unchanged. The simplest rephasing invariant constructed from the CKM matrix is the Jarlskog invariant~$J$, defined by $\im(V_{ij}V^{-1}_{jk}V_{k\ell}V^{-1}_{\ell i})=J\epsilon_{ik}\epsilon_{j\ell}$. Any CP violating bosonic operator thus has to contain at least four $W^\pm$ fields. As shown by Smit~\cite{smit}, there is, however, no CP violation even at fourth order of the gradient expansion at zero temperature---a conclusion straightforwardly generalizable to nonzero temperature. This in particular implies that the anomalous Wess-Zumino-Witten term need not be considered, as it only contributes at fourth order~\cite{salcedo6}. Hence, the CP violating part of the bosonic effective action can be determined solely using the Dirac operators (\ref{dirac1}) and (\ref{dirac2}) as well as Eqs.~(\ref{gammaplus}) and (\ref{gammaminus}).

Finally, we note that the above exercise can naturally be repeated for the lepton sector of the SM, assuming the neutrinos to have Dirac type masses. Their contribution is, however, always suppressed by many orders of magnitude with respect to the quarks, and we have therefore not considered it in our work.

%%%%%%%%%%%%%%%%%%%%%%%%%%%%%%%%%%%%%%%%%%%%%%%%%%%%%%%%%%%%%%%%%%%%%%%%

\emph{Results}.---We have implemented the gradient expansion and performed the traces indicated in Eqs.~(\ref{gammaplus}) and (\ref{gammaminus}) using the Mathematica package Feyncalc \cite{Mertig:1990an}. At the leading non-vanishing (sixth) order, the CP violating part of the effective action can be written in the form
\begin{eqnarray}
&&\Gamma_{\text{CP-odd}}= \label{action1}\\
&&-\frac\imag2N_cJG_{\text{F}}\kappa_{\text{CP}}\int_0^{1/T} dx_0\int d^3\vek x\left(\frac v\phi\right)^2({\mathcal O}_0+{\mathcal O}_1+{\mathcal O}_2),\nonumber
\end{eqnarray}
where $N_c=3$ is the number of colors, $G_{\text{F}}=1/(\sqrt2v^2)$ the Fermi coupling, and the zero temperature effective coupling $\kappa_{\text{CP}}$ reads
\begin{eqnarray}
\kappa_{\text{CP}}&\equiv&\frac{\Delta}{G_{\text{F}}}\int\frac{d^4p}{(2\pi)^4}(p^2)^3\prod_{f=1}^6\frac1{(p^2+m_f^2)^2}\approx3.1\times10^2,\nonumber\\
\Delta&\equiv&(m_u^2-m_c^2)(m_u^2-m_t^2)(m_c^2-m_t^2)\nonumber\\
&&\times(m_d^2-m_s^2)(m_d^2-m_b^2)(m_s^2-m_b^2).
\end{eqnarray}
The index $n$ in the functions ${\mathcal O}_n$ counts the number of $Z$ or $\varphi$ fields where $\varphi_\mu\equiv(\partial_\mu\phi)/\phi$. Each of these three terms can be subsequently divided into P even and P odd parts, $\mathcal O_n=\mathcal O^+_n+\mathcal O^-_n$. These functions contain not only Lorentz invariant operators, but also ones containing one or several vectors $u_\mu\equiv \delta_{\mu0}$, specifying the rest frame of the thermal bath. The non-invariant terms must naturally vanish in the limit of zero temperature, which we have indeed verified.

The Lorentz invariant operators in $\mathcal O^+_n$ read explicitly
\begin{eqnarray}
{\mathcal O}_0^+
&=&-\frac{c_1}{3}(W^+)^2W^-_{\mu\mu}W^-_{\nu\nu}
+\frac{5c_2}{3}(W^+)^2W^-_{\mu\nu}W^-_{\mu\nu}\nonumber\\
&&-\frac{c_1}{3}(W^+)^2W^-_{\mu\nu}W^-_{\nu\mu}
+\frac{4c_3}{3}W^+_\mu W^+_\nu W^-_{\mu\alpha}W^-_{\alpha\nu}\nonumber\\
&&-\frac{2c_1}{3}W^+_\mu W^+_\nu W^-_{\mu\alpha}W^-_{\nu\alpha}
-2c_4W^+_\mu W^+_\nu W^-_{\alpha\mu}W^-_{\alpha\nu}\nonumber\\
&&+\frac{4c_3}{3}W^+_\mu W^+_\nu W^-_{\mu\nu}W^-_{\alpha\alpha}-\cc ,\\
{\mathcal O}_1^+&=&
\frac83(Z_\mu+\varphi_\mu)\bigl[c_5(W^+)^2W^-_\mu W^-_{\nu\nu}\nonumber \\
&&-c_6(W^+)^2W^-_\nu W^-_{\mu\nu}
-c_6(W^+)^2W^-_\nu W^-_{\nu\mu}\nonumber\\
&&-c_3(W^+\cdot W^-)W^+_\mu W^-_{\nu\nu}\nonumber \\
&&+c_7(W^+\cdot W^-)W^+_\nu W^-_{\mu\nu}
+c_7W^+_\mu W^+_\nu W^-_\alpha W^-_{\alpha\nu}\nonumber\\
&&-c_{12}(W^+\cdot W^-) W^+_\nu W^-_{\nu\mu}
 -c_{12} W^+_\mu W^+_\nu W^-_\alpha W^-_{\nu\alpha}\nonumber\\
 &&+c_{13} W^-_\mu W^+_\nu W^+_\alpha W^-_{\nu\alpha}\bigr]-\cc ,\\
{\mathcal O}_2^+&=&
4(Z_\mu Z_\nu+\varphi_\mu\varphi_\nu)\nonumber\\
&&\times\bigl[c_8(W^+)^2W^-_\mu W^-_\nu
-c_8(W^-)^2W^+_\mu W^+_\nu\bigr]\nonumber\\
&&-\frac{16}3(Z\cdot\varphi)\bigl[c_9(W^+\cdot W^-)^2-2c_6(W^+)^2(W^-)^2\bigr]\nonumber\\
&&+\frac43(Z_\mu\varphi_\nu+Z_\nu\varphi_\mu)\nonumber\\
&&\times\bigl[c_{10}(W^+)^2W^-_\mu W^-_\nu+c_{10}(W^-)^2W^+_\mu W^+_\nu\nonumber\\
&&-2c_{11}(W^+\cdot W^-)(W^+_\mu W^-_\nu+W^+_\nu W^-_\mu)\bigr],
\label{action2}
\end{eqnarray}
while we find that
\begin{eqnarray}
{\mathcal O}_0^-={\mathcal O}_1^-={\mathcal O}_2^-=0.
\end{eqnarray}
Here, ``$\cc$'' stands for complex conjugation, acting on the fields as $W^\pm\to-W^\mp$, $Z\to-Z$, and $\varphi\to\varphi$. In addition, we have denoted the hypercharge covariant derivatives of $W^\pm$ by $W^\pm_{\mu\nu}\equiv(\partial_\mu\pm B_\mu)W^\pm_\nu$. For the numerical values of the quark masses and the Jarlskog invariant (see Ref.~\cite{PDG}), we have used $m_u=2.5\text{ MeV}$, $m_d=5\text{ MeV}$, $m_c=1.27\text{ GeV}$, $m_s=100\text{ MeV}$, $m_t=172\text{ GeV}$, $m_b=4.2\text{ GeV}$ as well as $J=2.9\times 10^{-5}$.

%%%%%%%%%%%%%%%%%%%%%%%%%%%%%%%%%%%%%%%%%%%%%%%%%%%%%%%%%%%%%%%%%%%%%%%%
\begin{figure}[t]
\includegraphics[width=\columnwidth]{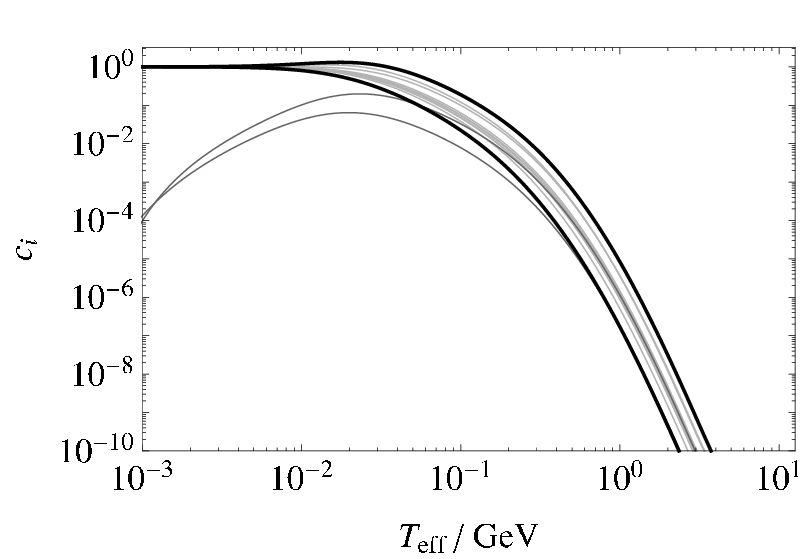}
\caption{The coefficients $c_1$--$c_{13}$ plotted as functions of the effective temperature $T_\text{eff}\equiv vT/\phi$. The bold lines correspond to the smallest and largest of the $c_i$ that approach one at zero temperature, i.e.~$c_1$ and $c_{10}$. 
}
\label{fig1}
\end{figure}
%%%%%%%%%%%%%%%%%%%%%%%%%%%%%%%%%%%%%%%%%%%%%%%%%%%%%%%%%%%%%%%%%%%%%%%%

The coefficients $c_i$ depend on the quark masses $m_f$ as well as the temperature $T$ and the Higgs field $\phi(x)$, which appear in the particular combination $T_{\text{eff}}\equiv vT/\phi$. In the zero temperature limit, $c_1$--$c_{11}$ approach unity while $c_{12}$--$c_{13}$ tend to zero, reducing our result to that of Ref.~\cite{salcedo6} and thus independently verifying its conclusions.

In Fig.~\ref{fig1}, we display the behavior of the coefficients $c_i$ as functions of $T_\text{eff}$, obtained through numerical evaluation of their defining one-loop, multi scale sum-integrals. At low temperatures, the $c_1$--$c_{11}$ evolve slowly until at $T_\text{eff}\simeq10$--$30\text{ MeV}$ they begin to fall rapidly. The $c_{12}$--$c_{13}$ on the other hand first exhibit a fast increase, reaching their maximum around $20$--$30$ MeV, after which they, too, begin to decrease. The largest coefficient at all temperatures is $c_{10}$, which reaches its maximal value of $1.3$ at $T_\text{eff}\approx 18\text{ MeV}$. 
In addition, we note that at $T_\text{eff}\gtrsim 10$ GeV, all $c_i$'s are at most of order $10^{-14}$, consistent with the estimates of Shaposhnikov \emph{et al.}~\cite{shaposhnikov}.

%%%%%%%%%%%%%%%%%%%%%%%%%%%%%%%%%%%%%%%%%%%%%%%%%%%%%%%%%%%%%%%%%%%%%%%%
\begin{figure}[t]
\includegraphics[width=\columnwidth]{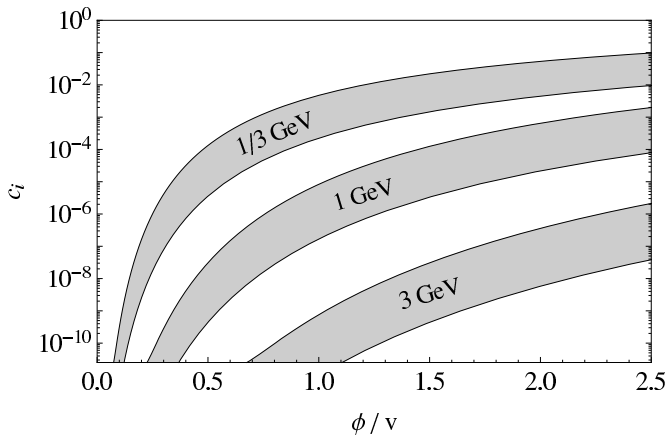}
\caption{The dependence of the $c_i$ on the Higgs field $\phi$, plotted for three different temperatures, $T=1/3\text{ GeV}$, $1\text{ GeV}$ and $3\text{ GeV}$. In each case, the grey band is spanned by $c_1$ and $c_{10}$.}
\label{fig2}
\end{figure}
%%%%%%%%%%%%%%%%%%%%%%%%%%%%%%%%%%%%%%%%%%%%%%%%%%%%%%%%%%%%%%%%%%%%%%%%

Fig.~\ref{fig2}, on the other hand, demonstrates the dependence of our results on the value of the Higgs field $\phi$. As is evident from the functional form of $T_{\text{eff}}=vT/\phi$, larger values of $\phi$ ameliorate the thermal suppression of CP violating effects. This observation has important implications for cold electroweak baryogenesis simulations, highlighting the necessity of determining the distribution of the Higgs field during the cold spinodal transition \cite{vanderMeulen:2005sp}.

Finally, we want to stress that in the present work we have completely neglected the effects of the strong interaction, an issue of increasing severity as one approaches the deconfinement transition (see also the discussion in Ref.~\cite{salcedo8}).  For an extended discussion of this issue as well as of the details of our computation (including the Lorentz breaking operators present at nonzero temperature), we refer the reader to Ref.~\cite{longpaper}.

%%%%%%%%%%%%%%%%%%%%%%%%%%%%%%%%%%%%%%%%%%%%%%%%%%%%%%%%%%%%%%%%%%%%%%%%

\emph{Conclusions and outlook}.---In this Letter, we have determined the leading CP violating operators in the covariant gradient expansion of the bosonic effective action of the Standard Model, displayed in Eqs.~(\ref{action1})--(\ref{action2}) above. In the zero temperature limit, we independently confirmed the result of Ref.~\cite{salcedo6}. While this agreement itself does not resolve the conflict between the results of Refs.~\cite{schmidt,salcedo6}, it suggests that the source of the discrepancy must lie either in the details of the computation performed in Ref.~\cite{schmidt}, or in a subtlety in separating $\Gamma^-$ into an anomalous and a regular part, cf.~Eq.~(\ref{gammaminus}). 

The gradient expansion we employ involves setting the external momenta to zero. One should bear in mind that this may limit the range of applicability of the result. In agreement with Refs.~\cite{schmidt, salcedo6}, we expect the gradient expansion to be valid up to external momenta of the order of $m_c$. This is based on the observation that the result is insensitive to the three smallest quark masses. In Ref.~\cite{longpaper}, we address the issue of convergence by calculating the next order in the expansion.

Nonzero temperature was observed to affect the values of the couplings of the operators already present at zero temperature, as well as to give rise to altogether new operators. 
Still, similarly to the zero temperature limit,
 all CP violating operators turned out to be P even/C odd. 
 Another important feature of our result was that the temperature entered only through the combination $T_\text{eff}\equiv vT/\phi$, involving the local Higgs field $\phi$ and its vacuum expectation value~$v$.

Direct numerical simulations using the SM effective action with 
a P-odd operator (of the ${\mathcal O}_1^-$-type in our classification) suggest that the cold electroweak baryogenesis scenario is able to produce the observed baryon asymmetry of the Universe with $c_{i}\simeq 10^{-5}$~\cite{tranbergshort,tranberglong}. During the phase transition, the fields are strongly out of equilibrium, and thus one would ideally like to derive the CP violating operators in such an environment. As this task is, however, technically extremely demanding (unless one includes the fermions themselves in the dynamics, which has become viable only recently~\cite{saffin}), our calculation assumed thermal equilibrium at an effective temperature $T$. As a rough estimate of the temperature range, in which cold electroweak baryogenesis might be viable, we observe that for $\phi=v$, the largest $c_i$ coefficient $c_{10}$ reaches the value $10^{-5}$ at $T\simeq 1$ GeV. It should be noted, however, that the effective action of Eqs.~(\ref{action1})--(\ref{action2}) consists of a variety of operators with coefficients of different magnitude, which may either enhance each others' effects or lead to surprising cancelations.

The result we have obtained should be contrasted with the energy 
scales typically associated with cold electroweak baryogenesis. If the relevant momentum scale of the dynamics is identified with the inverse time scale of the electroweak transition, lattice simulations using a hypothetical order four operator indicate an effective temperature of $T\simeq \tau_Q^{-1}\simeq5$--$10\text{ GeV}$, where $\tau_Q$ is the maximal time scale of a transition sufficiently out of equilibrium to generate a large enough baryon asymmetry~\cite{tranbergspeed}. 
These numbers are somewhat, but not dramatically higher than the approximative 1 GeV we obtained. To further quantify the result, classical numerical simulations of the bosonic effective action, including all of the operators we derived, are clearly required. 

An important technical note in this context is that, since in the bosonic simulations the asymmetry is quantified via the Chern-Simons number rather than the baryon number itself, the dynamics have to be explicitly CP but also P violating to generate an asymmetry. Given that the CP violating terms found here conserve P, one must in addition include the C odd/P odd but CP even operators. These are inherent in the fermionic sector of the SM and appear at lower orders in the derivative expansion of the bosonic effective action than the CP violating operators. We will further address this important point in Ref.~\cite{longpaper}.
Beyond this, improving the accuracy of the predictions would require not integrating out the fermions but rather incorporating them into the (quantum) simulations. We plan to return to these questions in future work.

%%%%%%%%%%%%%%%%%%%%%%%%%%%%%%%%%%%%%%%%%%%%%%%%%%%%%%%%%%%%%%%%%%%%%%%%

\begin{acknowledgments}
We are grateful to Lorenzo L.~Salcedo for his insightful comments, and to J\"urgen Berges, Thomas Konstandin, Mikko Laine and Jan Smit for useful discussions. T.B.,~O.T.~and A.V.~were supported by the Sofja Kovalevskaja program of the Alexander von Humboldt Foundation, and A.T.~by the Carlsberg Foundation.
\end{acknowledgments}

%%%%%%%%%%%%%%%%%%%%%%%%%%%%%%%%%%%%%%%%%%%%%%%%%%%%%%%%%%%%%%%%%%%%%%%%

\end{document}